# Irrelevance of magnetic proximity effect to spin-orbit torques in heavy metal/ferromagnet bilayers

L. J. Zhu,[1]* D. C. Ralph,[1,2] and R. A. Buhrman[1]
1. Cornell University, Ithaca, NY 14850
2. Kavli Institute at Cornell, New York, NY 14853, USA
*e-mail: lz442@cornell.edu

The magnetic proximity effect (MPE) is a well-established magnetic phenomenon that occurs at certain heavy metal (HM)/ferromagnet (FM) interfaces. However, there is still an active debate as to whether the presence of a MPE affects spin transport through such a HM/FM interface. Here we demonstrate that the MPE at Pt/Co and Au$_{0.25}$Pt$_{0.75}$/Co interfaces can be enhanced substantially by thermal annealing protocols. From this ability, we show that the MPE has no discernable influence on either the damping-like or the field-like spin-orbit torques exerted on the FM layer due to the spin Hall effect of the HM layer, indicating a minimal role of the MPE compared to other interfacial effects, e.g. spin memory loss and spin backflow.

**Key words**: Spin orbit torque, Spin Hall effect, Proximity effect, Spin orbit coupling

The magnetic proximity effect (MPE) is an interfacial magnetic phenomenon whereby a ferromagnetic (FM) layer induces a magnetic moment in a neighboring heavy metal (HM) or semiconductor [1,2] due to an exchange interaction that decays rapidly away from the interface (see Fig. 1(a)). Despite intensive theoretical and experimental efforts [3-8], it has remained in debate as to whether a strong MPE at a HM/FM interface has a significant effect on the spin transparency of the interface ($T_{\text{int}}$) and hence on the spin-orbit torques (SOTs) exerted on the FM layer by spin currents from the spin Hall effect (SHE) of the HM layer. Degradation of $T_{\text{int}}$ is known to occur due to other interfacial effects, namely spin backflow (SBF)[9] and spin memory loss (SML) [10-16]. In regard to the MPE, however, so far it has been reported as suppressing [3,4], enhancing [7,8], or being irrelevant to [10] interfacial spin mixing conductance ($G^{\uparrow\downarrow}$) and thus to $T_{\text{int}}$ and SOT efficiencies. Recently the MPE at the Pt/Co$_2$FeAl interface was suggested by Peterson et al. [6] to be irrelevant to the damping-like SOT efficiency ($\xi_{\text{DL}}$) while suppressing the field-like SOT efficiency ($\xi_{\text{FL}}$) at low temperatures where the MPE was argued to be the strongest. Part of the reason why the role of the MPE remains unsettled is that it is challenging to significantly vary the strength of the MPE in a given bilayer material system while maintaining SOTs strong enough to determine accurately.

In this Rapid Communication, we report that the strength of the MPE at Pt/Co and Au$_{0.25}$Pt$_{0.75}$/Co interfaces, where strong SOTs arise from the SHE of the HMs [17], can be tuned significantly by varying thermal annealing conditions. From this ability we obtain through direct SOT measurements evidence that there is no discernable correlation between the strength of the MPE and the SOT efficiencies resulting from the SHE in the HMs.

As listed in Table I, the magnetic stacks for this work are comprised of three sample series: Pt 4/Co 0.85 (samples P1-P4) and Au$_{0.25}$Pt$_{0.75}$ 4/Co 0.85 (samples P5-P8) with perpendicular magnetic anisotropy (PMA) (the numbers are layer thicknesses in nm) and Pt 4 /Hf 0.67/ Co 1.4 (samples R1-R3) with in-plane magnetic anisotropy (IMA). Each stack was deposited by DC/RF sputtering onto oxidized Si substrates with a 1 nm Ta seed layer, and capped by a 2 nm MgO layer and finally a 1.5 nm Ta layer that was fully oxidized upon exposure to atmosphere [17]. Each layer was sputtered at a low rate (e.g. 0.007 nm/s for Co and 0.016 nm/s for Pt) by introducing an oblique orientation of the target to the substrate and by using low magnetron sputter power to minimize intermixing. Each stack underwent repeated cycles of measurements and annealing to tune the strengths of the proximity magnetism. X-ray diffraction (XRD) measurements indicate that the HM and Co layers are textured with a (111) normal orientation [18] as is usually found in these types of multilayers [11,19,20]. Annealing was performed in a vacuum furnace with a background pressure of ~10$^{-7}$ Torr. The magnetic moment of a 0.50×0.46 cm$^2$ piece of each sample (~10$^{-5}$-10$^{-4}$ emu) was measured at 300 K with a standard VSM (sensitivity ~10$^{-7}$ emu) embedded in a Quantum Design physical property measurement system. The samples were further patterned into 5×60 μm$^2$ Hall bars for SOT studies.

We first show that a MPE is likely present at the as-grown Pt/Co and Au$_{0.25}$Pt$_{0.75}$/Co interfaces and, in any case, can be significantly enhanced simply by annealing. As an example, Fig. 1(b) plots the measured effective magnetization ($M^{\text{eff}}$), determined assuming that the measured magnetic moment is contributed solely by the Co layer with the deposited thickness, as a function of in-plane magnetic field for the Pt/Co sample set (P1-P4). An enhancement of the total moment due to the thermal annealing can be clearly seen. In Fig. 2(a) we summarize the saturation values of $M^{\text{eff}}$, i.e., $M_s^{\text{eff}}$, for the different samples. For the PMA samples, $M_s^{\text{eff}}$ is gradually and monotonically enhanced from ~1510 (1410) emu/cm$^3$ in P1 (P5) to ~1970 (2050) emu/cm$^3$ in P4 (P8). Interestingly, $M_s^{\text{eff}}$ in P1-P4, P7, and P8 is, in all cases, apparently larger than that of Co bulk (~1450 emu/cm$^3$ [21] or ~1.74 μ$_B$/Co), the value marked with a blue dashed line in Fig. 2(a).

We can reasonably exclude inter-mixing and alloying at the interface as the cause of the large increases in $M_s^{\text{eff}}$ for the Pt/Co and Au$_{0.25}$Pt$_{0.75}$/Co bilayers upon annealing. First, chemically disordered Co-Pt alloys that are Pt-rich (85% Pt) and have the fcc phase (A1) are reported to be paramagnetic at 300 K [22]. Second, *chemically ordered* Pt-rich ferromagnetic mixtures (e.g. $L1_2$–CoPt$_3$) have Curie temperatures of < 300 K [22]. Therefore, an interfacial region of Co intermixed into Pt forming either the chemically disordered A1 or the ordered $L1_2$–CoPt$_3$ phase should result in a magnetic dead layer at room temperature [11] and therefore a reduction in $M_s^{\text{eff}}$. This does not exclude of course the possibility that such a paramagnetic layer of A1 Co-Pt mixture at the interface (including any possible grain boundary areas and crystalline defects, e.g. threading dislocation) is part of the material that becomes magnetic at room temperature due to the proximity effect from the adjacent Co (rich) layer [2]. For Pt intermixed into Co, which is less likely given the deposition order (i.e. Si/SiO$_2$/Ta 1/Pt or Au$_{0.25}$Pt$_{0.75}$ 4/Co 0.85/MgO 2/Ta 1.5), while the Curie point can be above room temperature, if the alloy is disordered $M_s$ would be low. For example, in sputtered thin films of CoPt $M_s$ was found to be <300 emu/cm$^3$ (~0.86 µ$_B$/Co) due to chemical disorder [23]. Finally, if the annealing process resulted in the formation of an interfacial layer of *highly chemically ordered* material of either the $L1_0$-CoPt or $L1_2$–Co$_3$Pt phase, the upper limit for the saturation magnetization of that layer would be the bulk values of $M_s$ ≈ 740 emu/cm$^3$ for $L1_0$-CoPt (~2.13 µ$_B$/Co, lattice constant $a$ = 3.803 Å, $c$ = 3.701 Å)[24] and $M_s$ ≈ 970 emu/cm$^3$ for $L1_2$–Co$_3$Pt (~1.72 µ$_B$/Co, $a$ = 3.668 Å)[25]. Apparently, the moment per Co atom in $L1_2$–Co$_3$Pt is smaller than in Co bulk and thus than that measured in our samples. The effective Co moment in $L1_0$-CoPt, where Co and Pt atomic layers stack alternatively along the $c$ axis, is large due to the proximity of Pt atoms to the magnetic Co atoms in the sharp Pt/Co superlattices (there is no Pt in the Co layer and no Co atoms in the Pt layer). Even if all of the Co atoms were incorporated in a highly chemically ordered layer of $L1_0$-CoPt, the result would be an upper limit $M_s^{\text{eff}}$ of 1775 emu/cm$^3$ (~2.13 µ$_B$/Co), which still fails to explain the large $M_s^{\text{eff}}$ of 1950 to 2050 emu/cm$^3$ observed in samples P4 and P8. Thus there must be a longer scale for the proximity effect than only one magnetic Pt atomic layer for one Co layer. An additional line of evidence that safely excludes a formation of $L1_0$-CoPt is that the magnetic easy axis of $L1_0$-CoPt is along (001) crystalline direction [26], which means that a gradual magnetic hysteresis loop would be observed in the $L1_0$-CoPt (111) direction. This is contrary to the observation of fairly sharp magnetization switching driven by an out-of-plane field in all the as-grown and annealed Pt/Co (111) or Au$_{0.25}$Pt$_{0.75}$/Co (111) bilayer samples (see below). We also note that the PMA in our samples is strong upon deposition and improves upon annealing [18], with the interfacial magnetic anisotropy energy density $K_s$ ≈ 1-3.5 erg/cm$^3$. This indicates that the interface is becoming less intermixed and more ordered as the result, since chemical disorder is expected to degrade the PMA [27,28]. This observation is also supported by the enhanced oscillation and attenuation length of the x-ray reflectivity of the bilayers after annealing [18].

Relaxation of elastic strain at the HM/Co as the result of annealing does not provide a ready explanation for the significant enhancement of $M_s^{\text{eff}}$ that we observe. The XRD results indicate that there is a 0.6% and 1.3% increase of the in-plane lattice constant for the P1-P4 and P5-P8 HM layers, respectively [18], which, even if fully mirrored by the Co layer seems too small to make any significant change in the band structure and therefore in the Co magnetization. Note that Lee *et al*. found that a large strain of up to 2% has no distinguishable influence on the magnetization of Co [29]. The absence of a significant correlation between $M_s^{\text{eff}}$ and any interface-generated Co strain is also indicated by the independence of the Co magnetization and the proximity moment on the Co thickness when that was varied from 0.85 to 1.1 nm, since we would expect strain due to lattice mismatch to be more relaxed in the thicker Co [18]. The irrelevance of this film strain to the magnetization (and magneto-optical Kerr angles) of Co layers has been also well established in Pt/Co or Au/Co multilayers [20,30].

As result of the above considerations we attribute the observed significant enhancement in the total magnetic moment or $M_s^{\text{eff}}$ to a strong MPE at the HM/Co interface, due to which the first few HM atomic layers adjacent to Co become magnetized, and that the strength of this MPE increases monotonically with annealing. The $M_s^{\text{eff}}$ values considerably larger than that of Co bulk are consistent with previous reports in unannealed Pt/Co supperlattices with pronounced MPE, e.g. 2250 emu/cm$^3$ [21] or 2000-3000 emu/cm$^3$ [31] in Pt/Co multilayers. In those Pt/Co systems, the strongest MPE was found when Co was only one atomic layer thick [21,31]. As an additional check, a passivating spacer with a low Stoner factor, e.g. Hf, should suppress the MPE between Co and Pt. Consistent with that assumption we found that, as can be seen in Fig. 2(a), as-grown Pt 4/ Hf 0.67/ Co 1.4 has $M_s^{\text{eff}}$ ≈ 1330 emu/cm$^3$, which remains unchanged after annealing at 300 ºC for 0.5 h (R2) and 350 ºC for 2h (R3).

To give a better sense of the MPE strength and of its variation upon annealing, in Fig. 2(b) we determined the proximity-induced magnetic moment per unit area ($\Delta M/A$) of the samples by assuming a constant $M_s$ of 1330 emu/cm$^3$ for the Co thin films, the same value as in Pt 4/Hf 0.67/ Co 1.4 system. Despite that the MPE induced $M_s$ of the magnetized HM decays away from the interface, we introduce in Fig. 2(c) an "effective" thickness $\Delta t_{\text{eff}}$ for the magnetized HM layer, i.e. $\Delta t_{\text{eff}} = \Delta M/M_s(\text{HM})$ to account for the MPE contribution with the assumption of a depth-independent $M_s(\text{HM})$ of 420 emu/cm$^3$, i.e. ~0.68 µ$_B$/Pt, as determined by element-specific x-ray magnetic circular dichroism (XMCD) measurements on as-grown Pt/Co multilayers where the Pt layer was 3 atomic layers thick [32]. In this representation $\Delta t_{\text{eff}}$ increases monotonically from 0.37 to 1.28 nm for the Pt/Co interface and from 0.17 to 1.45 nm for the Au$_{0.25}$Pt$_{0.75}$/Co interface. The $\Delta t_{\text{eff}}$ values in the as-grown samples are fairly consistent with those in previous XMCD studies on un-annealed Pt/Co multilayers [33], while $\Delta t_{\text{eff}}$ after the final annealing step is considerably larger. This representation of the MPE is just for illustrative purposes, because the likely microscopic situation is that the annealing is enhancing the average exchange interaction at the HM/Co interface by improving interfacial order, and hence increasing the average induced moment on the interfacial Pt, rather than changing the MPE decay length. An XMCD study of Pt/Co multilayers as a function of annealing could be informative for understanding the MPE mechanism in detail, but, to the best of our knowledge, such a study has not yet been pursued.

We now turn to discuss the behavior of the SOTs in Pt/Co and Au$_{0.25}$Pt$_{0.75}$/Co bilayers. We determined SOT efficiencies for the PMA HM/Co bilayers by harmonic Hall measurements [11,16,17] with a 4 V excitation applied to the current leads of the Hall bar which is along $x$ direction (see Fig. 3(a)). As noted above all the Pt/Co and Au$_{0.25}$Pt$_{0.75}$/Co bilayers show strong PMA as indicated by the fairly square anomalous Hall voltage hysteresis loops (see Fig. 3(b)). The damping-like (field-like) effective spin torque fields are given by $H_{\text{DL(FL)}}$ = $-2\frac{\partial V_{2\omega}}{\partial H_{x(y)}}/\frac{\partial^2 V_{1\omega}}{\partial^2 H_{x(y)}}$, where the first and second harmonic Hall voltages, $V_{1\omega}$ and $V_{2\omega}$, are parabolic and linear functions of in-plane bias fields $H_x$ and $H_y$ (see Figs. 3(c) and 3(d)), respectively. In Fig. 4, we show damping-like and field-like SOT efficiencies for the samples before and after annealing as determined following $\xi_{DL(FL)}^j = 2e\mu_0 M_s^{eff} t H_{DL(FL)}/\hbar j_e$, with $e$, $\mu_0$, $t$, $\hbar$, and $j_e$ being the elementary charge, the permeability of vacuum, the ferromagnetic layer thickness, the reduced Planck constant, and the charge current density, respectively. For both the Pt/Co series (P1-P4) and Au$_{0.25}$Pt$_{0.75}$/Co series (P5-P8), upon the first annealing step, $\xi_{DL}^j$ and $\xi_{FL}^j$ consistently drop by ~50% in magnitude and then gradually recover back to some extent as a result of the two subsequent annealing steps. Obviously, the variations of $\xi_{DL(FL)}^j$ upon annealing (Fig. 4) are distinctly different from those of MPE characterized by $M_s^{eff}$ and $\Delta t_{eff}$ (Fig. 2), which safely excludes any important correlation of the MPE to $T_{int}$ or the spin transport from the HM into FM layer and thus the SOTs on the FM layer.

We should point out that the effect of annealing on the HM resistivity ($\rho_{HM}$) is minimal, with $\rho_{HM}$ increasing monotonically by just a small amount, $\rho_{Pt}$ ~57- 68 μΩ cm and $\rho_{AuPt}$ ~78- 81 μΩ cm [18]. The spin Hall conductivity $\sigma_{SH}$ for Pt and Au$_{1-x}$Pt$_x$ is dominated by the intrinsic SHE that is determined by the topology of the band structure [17], which for simple fcc metals is only dependent on the long-range crystal structure and hence is robust against localized changes in structural disorder that could occur during annealing (e.g. strain relaxation)[34]. The small increase of $\rho_{HM}$ is unlikely to be indicative of a significant change of $\sigma_{SH}$ of the Pt or Au$_{0.25}$Pt$_{0.75}$ layers, and thus changes in the spin Hall ratio ($\theta_{SH} = (2e/\hbar)\sigma_{SH}\rho_{HM}$) should be small. Since the spin diffusion length ($\lambda_s$) in these metals is understood to be set by the Elliott-Yafet spin relaxation mechanism ($\lambda_s \propto 1/\rho_{HM}$)[35-37], $\lambda_s$ would decrease *monotonically* by only a small amount with annealing, which cannot readily explain the non-monotonic variation of the spin-orbit torques with annealing. Instead, as we discuss elsewhere [16], we have found that the variation of the SOTs with annealing is a direct consequence of degradation of $T_{int}$ by the interfacial SOC that becomes stronger with annealing.

In light of these results we certainly need to consider other recent investigations of the possible role of the MPE in affecting interfacial spin transport. A spin pumping and first principles study by Zhang *et al.* reported a reduction or *loss* of the spin Hall conductivity in an ultrathin Pt layer (~0.6 nm) adjacent to a ferromagnetic NiFe layer due to the MPE [3]. This is in contrast to the conclusion of a first principles calculation by Guo *et al.* [38] that the MPE induced magnetic moment can slightly increase the spin Hall conductivity and anomalous Hall conductivity in Pt and Pd. However, our direct experimental data indicates that the proximity magnetism in Pt or Au$_{25}$Pt$_{75}$ has no distinguishable correlation with the strength of the spin-torques. We do note that in our case the Pt thickness, 4 nm, is larger than the effective thickness of the proximity layer, which is perhaps not the case in the studies by Zhang *et al.* [3] and Guo *et al.* [38].

In summary, we have demonstrated that annealing can substantially enhance the strength of MPE at Pt/Co and Au$_{25}$Pt$_{75}$/Co interfaces. This provides an experimental opportunity to determine that the MPE has minimal correlation with the efficiency of spin transport from the HM into the FM compared with other effects like interfacial spin-orbit scattering-induced SML, and therefore appears largely irrelevant to the magnitudes of the damping-like and field-like SOTs that are exerted on the FM layer. Our findings should be beneficial for better understanding of SOTs and MPE in HM/FM systems and their spintronic applications.

This work was supported in part by the Office of Naval Research and by the NSF MRSEC program (DMR-1719875) through the Cornell Center for Materials Research. Support was also provided by the Office of the Director of National Intelligence (ODNI), Intelligence Advanced Research Projects Activity (IARPA), via contract W911NF-14-C0089. The views and conclusions contained herein are those of the authors and should not be interpreted as necessarily representing the official policies or endorsements, either expressed or implied, of the ODNI, IARPA, or the U.S. Government. The U.S. Government is authorized to reproduce and distribute reprints for Governmental purposes notwithstanding any copyright annotation thereon. Additionally this work was supported by the NSF (ECCS-1542081) through use of the Cornell Nanofabrication Facility/National Nanotechnology Coordinated Infrastructure.

Table 1. Sample configurations and annealing conditions. Bilayers P1-P8 have perpendicular magnetic anisotropy, while R1-R3 have in-plane magnetic anisotropy. Layer thicknesses for Co, Pt, Au$_{0.25}$Pt$_{0.75}$ and Hf are in nm.

| # | Bilayers | Anneal condition |
|---|---|---|
| P1 | Pt 4/Co 0.85 | As-grown |
| P2 | Pt 4/Co 0.85 | 350 °C,2 h |
| P3 | Pt 4/Co 0.85 | 350 °C,4 h |
| P4 | Pt 4/Co 0.85 | 350 °C,4 h +400 °C,1 h |
| P5 | Au$_{0.25}$Pt$_{0.75}$ 4/Co 0.85 | As-grown |
| P6 | Au$_{0.25}$Pt$_{0.75}$ 4/Co 0.85 | 350 °C,2 h |
| P7 | Au$_{0.25}$Pt$_{0.75}$ 4/Co 0.85 | 350 °C,4 h |
| P8 | Au$_{0.25}$Pt$_{0.75}$ 4/Co 0.85 | 350 °C,4 h +400 °C,1 h |
| R1 | Pt 4/Hf 0.67/Co 1.4 | As-grown |
| R2 | Pt 4/Hf 0.67/Co 1.4 | 300 °C, 0.5 h |
| R3 | Pt 4/Hf 0.67/Co 1.4 | 350 °C, 2 h |

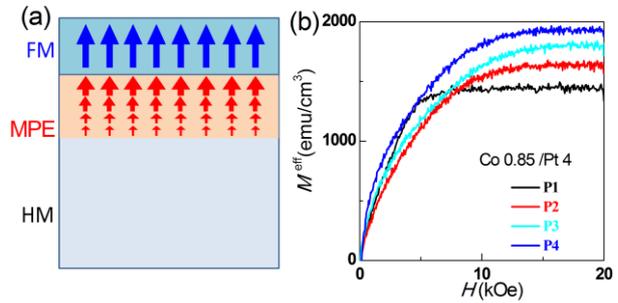

Fig. 1. (a) Schematic of magnetic proximity effect (MPE) at a FM/HM interface. (b) In-plane magnetization ($M$) at 300 K versus in-plane magnetic field ($H$) for a Pt 4/Co 0.85 bilayer annealed under different conditions (samples P1-P4).

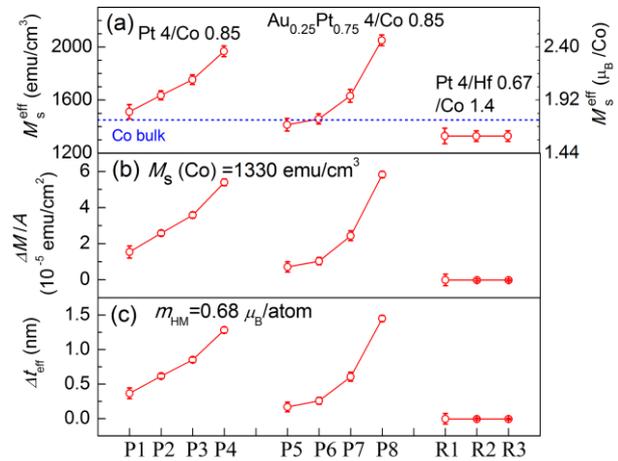

Fig. 2. (a) The effective saturation magnetization ($M_s^{eff}$) in the different sample series. (b) Proximity-induced additional magnetic moment per area obtained by subtracting the moment of the Co from the measured moment by assuming $M_s$(Co) = 1330 emu/cm$^3$. (c) Effective thickness of the induced magnetism in the HM assuming a uniform $M_s$ (HM) = 420 emu/cm$^3$ (0.68 $\mu_B$/atom). The blue dashed line in (a) denotes $M_s$ for Co bulk value of 1450 emu/cm$^3$.

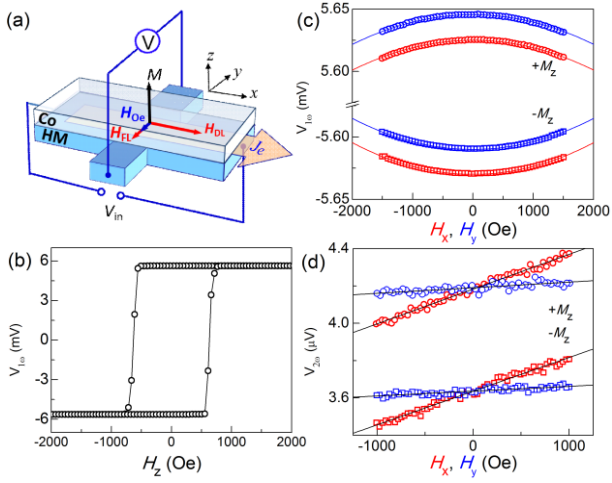

Fig. 3. (a) Schematic representation of the Hall bar device and measurement coordinates, (b) $V_{1\omega}$ versus $H_z$, (c) $V_{1\omega}$ versus $H_x$ (red) and $H_y$ (blue), (d) $V_{2\omega}$ versus $H_x$ (red) and $H_y$ (blue) for /Pt 4/Co 0.85 bilayer annealed at 350 °C for 4 h and 400 °C for 1 h (P4). In (c) and (d), top (bottom) two plots are for $+M_z$ ($-M_z$), and the solid lines refer to the best fits. In (c), the blue data points for $V_{1\omega}$-$H_y$ are artificially shifted by 0.02 mV for clarity.

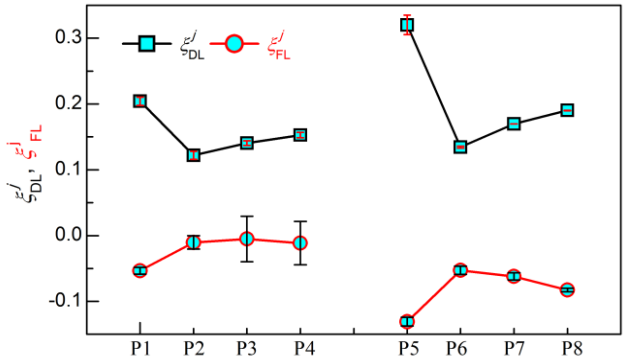

Fig. 4. Damping-like and field-like SOT efficiencies per unit bias current density for the sample series (P1-P8) defined in Table 1.

Supplemental Material

**Irrelevance of magnetic proximity effect to spin-orbit torques in heavy metal/ferromagnet bilayers**

L. J. Zhu, [1]* D. C. Ralph,[1,2] and R. A. Buhrman[1]
*1. Cornell University, Ithaca, NY 14850*
*2. Kavli Institute at Cornell, Ithaca, New York 14853, USA*

*e-mail: lz442@cornell.edu

**I. (111) orientation of heavy metal/Co bilayers**

**II. Annealing-induced strain variation in the bilayer samples**

**III. Resistivity of the heavy metal layers**

**IV. Enhancement of perpendicular magnetic anisotropy of a Pt 4/Co 1.44 bilayer**

# I. (111)-orientation of heavy metal/Co bilayers

X-ray diffraction measurements (see Fig. S1, Fig. S2, and Fig. S3) indicate that all the Pt/Co and $Au_{0.25}Pt_{0.75}$/Co stacks exhibit face-centered cubic structures with (111) orientation, in consistence with previous reports that a Ta buffer layer favors the stabilization of (111) orientated Pt [1]. Here we discuss the dominance of the (111) orientation of the HM/Co bilayer using the Pt 4/Co 0.85 bilayer as an example. For the (111), (100), and (110) orientations, the strongest allowed representative peaks are (111), (200), and (220) peaks, respectively. The Pt (111), (200) and (220) peaks, if any, should appear at 39.6º, 46.2º, and 67.5º, respectively. From a wide-range x-ray diffraction $\theta$-$2\theta$ scan, the orientation of the polycrystalline Pt/Co bilayer can be unambiguously determined. Figure S1 shows the wide-range x-ray diffraction $\theta$-$2\theta$ patterns of an as-grown Pt 4/Co 0.85 sample and that of a reference Si (001)/$SiO_2$ substrate. When the x-ray scattering vector is aligned along film normal (the black spectrum), the pattern only shows one peak from the Pt/Co bilayer, which is the (111) peak at 39.6º. In contrast, there is not any distinguishable (220) and (200) peak. In this case, only a weak Si (004) peak from the substrate is observed because of the miscut angle of the substrate. When the scattering vector is aligned along Si (001) direction (red), we observe several strong peaks from the Si/$SiO_2$ substrate, each of which overlaps very well with that of the bare Si/$SiO_2$ substrate (blue). Besides the substrate peaks, there is only a (111) peak from the Pt/Co bilayers. These observations clearly indicate that the Pt/Co system is highly (111) oriented, while there is a negligible amount material with any other orientation.

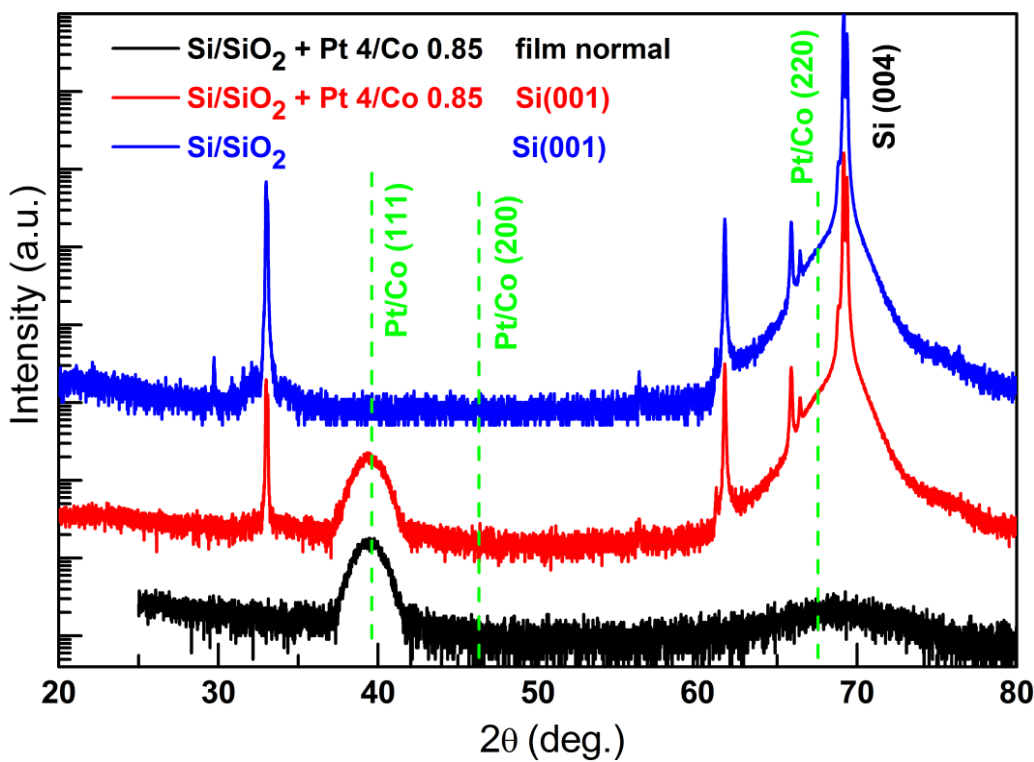

Fig. S1. Wide-range x-ray diffraction $\theta$-$2\theta$ patterns of an as-grown Pt 4/Co 0.85 sample and a bare Si (001)/$SiO_2$ substrate as a reference. The black and red patterns represent the spectra for the x-ray scattering vector aligned along the film normal and the Si (001) orientation, respectively. There is a 1º misalignment between the film normal and the Si (001) orientation due to miscut angle of the silicon substrate. The blue pattern is measured from a reference Si (001)/$SiO_2$ substrate with the x-ray scattering vector aligned along Si (001). The three green dashed lines mark the $2\theta$ positions of Pt/Co (111), (200), and (220) peaks, respectively.

## II. Annealing-induced strain variation in the bilayer samples

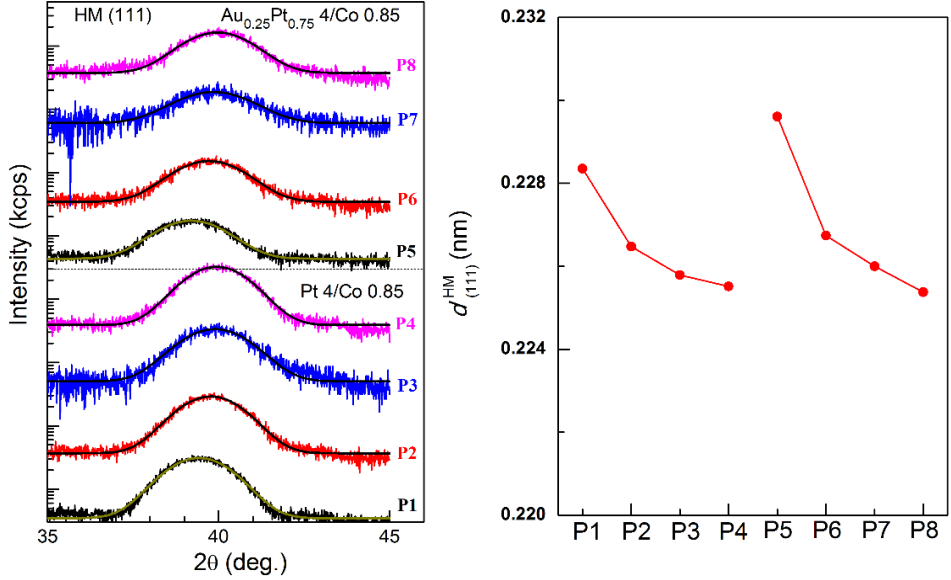

Fig. S2. Effect of annealing steps on the x-ray diffraction $\theta$-$2\theta$ patterns ($\chi = 0°$) for the sample series defined in Table 1 (left) and the indicated decrease with annealing of the out-of-plane (111) lattice constant of the Pt and $Au_{0.25}Pt_{0.75}$ HM layers (right).

In an x-ray diffraction $\theta$-$2\theta$ measurement, the peak position reflects the out-of-plane facet distance of the sample lattice. As shown in Fig. S2(a) and S2(b), The (111) peaks of these heavy metal (HM) /Co bilayers consistently shift to higher angles with annealing, which indicates a decrease in the (111) facet distances of the bilayers, i.e. $d_{(111)}$ = 0.228-0.226 nm for P1-P4 and 0.230-0.225 nm for P5-P8. It is very challenging to measure the in-plane lattice constants of these samples because the small layer thicknesses and the polycrystalline texture of the samples make the diffraction intensity for any diffraction peak direction off the film normal very weak, too weak to detect with our conventional high resolution thin-film X-ray diffractometer (see Fig. S3(a) and 3(b)). However, if we assume a minimal change of volume for the lattice cell due to strain, which is the usual situation, then the corresponding increase in the in-plane lattice constant of the HM layers can be estimated as $\Delta d_{(111)}/\sqrt{2}d_{(111)} \approx 0.6\%$ for P1-P4 and 1.3% for P5-P8, which even if fully mirrored by the Co layers due to a coherent strain. In our assessment this is too small to cause any significant change in the band structure and therefore in the magnetization of the Co layer. We note that the effective magnetization values of the bilayers after annealing are considerably *larger* (e.g. 2050 emu/cc in P8) than the Co bulk value (1450 emu/cc). It seems highly unlikely that such a 40% *increase* in the effective magnetization of Co can be due to such a small change (relaxation) of the as-grown strain in the Co layer. We have carried out an additional annealing experiment on a Pt 4/Co 3.2 bilayer sample where we directly find that the $d_{(111)}$ change for Co layer upon annealing is smaller than for the HM layers (see Fig. S4(a) and Fig. S4(b)). As shown in Fig. S5, there we also find a strong enhancement of magnetic moment similar to that of the two 0.85nm Co sample series (P1-P4 and P5-P8), which varies differently from the full width at half maximum (FWHM) of the (111) peaks that contains strain broadening contribution (see Fig. S4(c)). This strongly suggests that thin film strain, which should be dependent on Co thickness, cannot readily explain the experimental data, especially the absence of such enhancement of the magnetic moment in R1-R3 (see Fig. 2 in the main text), where a thin layer of Hf, a material with a low Stoner factor, is inserted between the Pt and Co layers.

We also measured the magnetic moments of an as-grown Pt 4/Co wedge sample as a function of Co thickness (strain), and find that the Co magnetization is independent of the thickness of Co layers, as indicated by the linear variation of the magnetic moment of the film with thickness shown in Fig. S5, at least for the thickness range studied (0.85 to 1.1 nm). This indicates that there is no significant correlation between the magnetization and any thickness dependent strain in the Co layers arising from the HM/Co interfaces. This is consistent with previous reports [2,3].

Roughness information of thin films can be determined from the incident angle-dependence studies of x-ray reflectivity. In x-ray reflectivity measurements, the oscillation strength and attenuation length represent the interface roughness. As can be seen from Fig. S3(d)-S3(d), the oscillation strength and attenuation length both indicate good interface sharpness in the as-grown state, and the quality of these signals is improved after the final-step annealing. This supports our conclusion that annealing-induced intermixing at the interfaces of the bilayer samples should be minimal and cannot readily explain the giant enhancement of the magnetic moment by annealing. However, a quantitative calculation of the roughness of each interface by fitting the overall x-ray reflectivity signals is not feasible here as there are too many layers (Si/SiO$_2$/Ta 1/Pt 4/Co 0.85/MgO 2/Ta 1.5) and too many parameters (densities, roughness, thicknesses, depth profile of each parameter, etc), to allow the nonlinear fit to reliably find the global minimum.

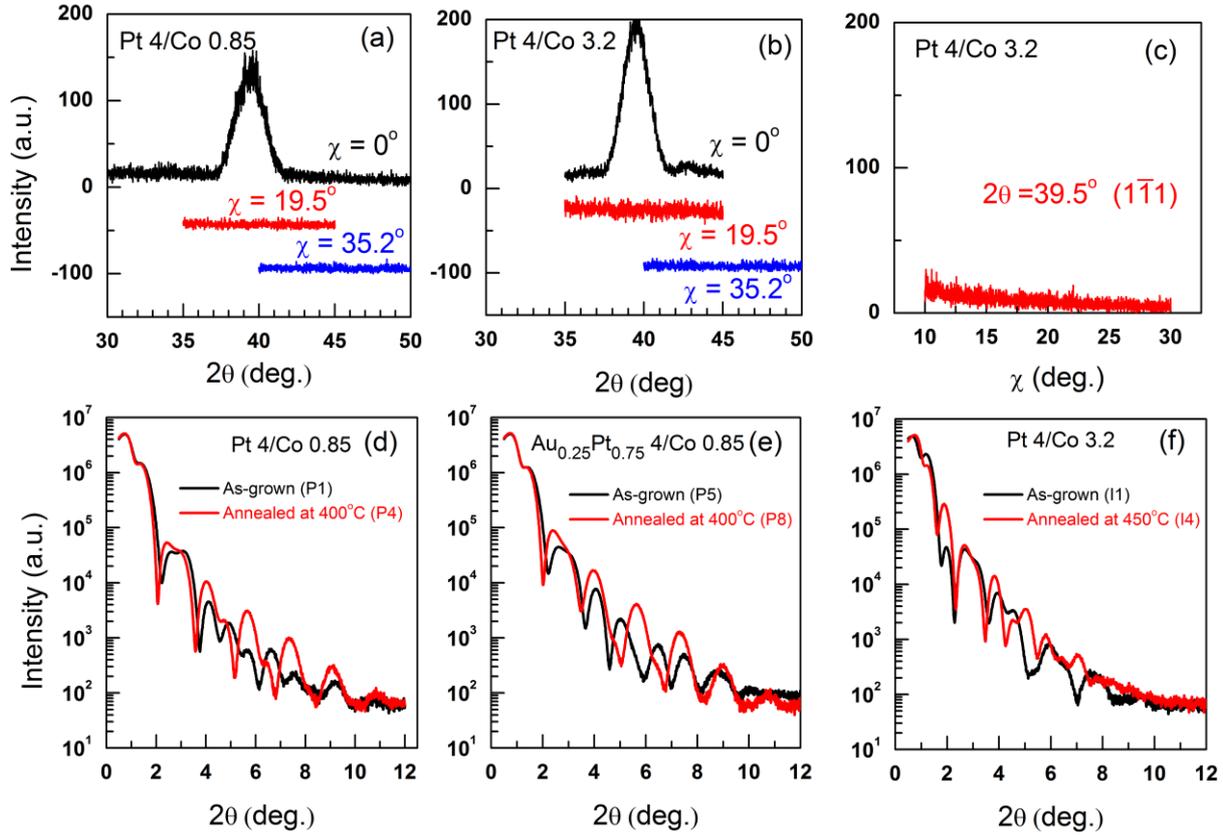

Fig. S3. X-ray diffraction $\theta$-$2\theta$ patterns at $\chi = 0°$, 19.5°, 35.2° for (a) as-grown Pt 4/Co 0.85, (b) as-grown Pt 4/Co 3.2; (c) $\chi$ scan at $2\theta \approx 39.5°$ for as-grown Pt 4/Co 3.2, indicating absence of the (1-11) or (-111) diffraction peaks, which are expected to appear at $\chi \approx 19.5°$. Here $\chi$ is the polar angle of the diffraction plane (containing the incident and reflected x-ray) away from film normal. Pt/Co (111) should have (111) diffraction peak at $2\theta \approx 39.5°$ for $\chi \approx 0°$, and (1-11) or (-111) diffraction peaks at $2\theta \approx 39.5°$ for $\chi \approx 19.5°$, and (200) diffraction peak at $2\theta \approx 46.2°$ for 35.2° if strong enough. There is no (1-11), (-111) or (200) observed in the polycrystalline samples due to the small layer thicknesses of Pt and Co and the averaging effect of the signal over the 360 degree due to the polycrystalline texture in the film plane. X-ray reflectivity curves for (d) as-grown (P1) and annealed (P4) Pt 4/Co 0.85; (e) as-grown (P5) and annealed (P8) Au$_{0.25}$Pt$_{0.75}$ 4/Co 0.85; (f) as-grown (I1) and annealed (I4) Pt 4/Co 3.2.

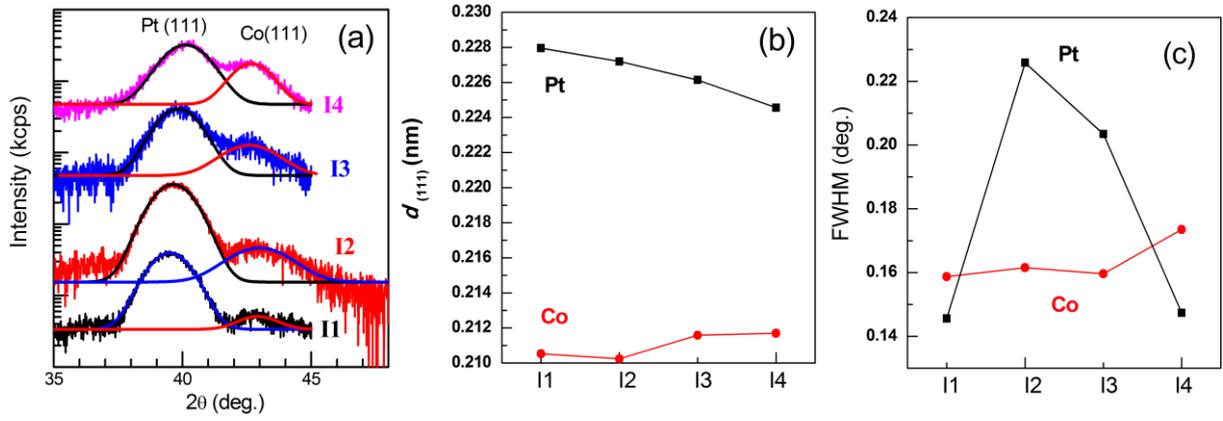

Fig. S4. (a) X-ray diffraction $\theta$-$2\theta$ patterns ($\chi = 0°$) for a Pt 4/Co 3.2 bilayer sample with enhanced annealing condition (I1-I4). The solid lines refer to best two-peak fits of the measured curves. (b) The facet distance d(111) for the Co and Pt layers. (c) Full width at half maximum of the (111) peaks of Co and Pt layers. The Pt (111) facet distance along film normal shows a similar change as in the PMA samples (P1-P8). However, the relative change of Co $d$ (111) is <0.5% from I1 to I4, which is much smaller than that of Pt. From the measured decrease of the out-of-plane lattice constant, the increase of the in-plane lattice constant of Co layer can be expected to be < 0.33%, a rather minimal change.

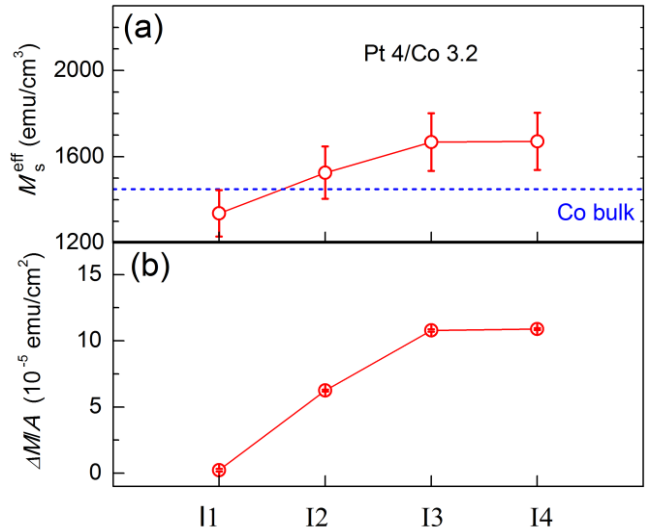

Fig. S5. (a) The effective saturation magnetization ($M_s^{\mathrm{eff}}$) and (b) Proximity-induced additional magnetic moment per area as the function of annealing for a Pt 4/Co 3.2 sample. The similar annealing steps were applied to this sample as to samples P1-P4 and P5-P8 as reported in the main text.

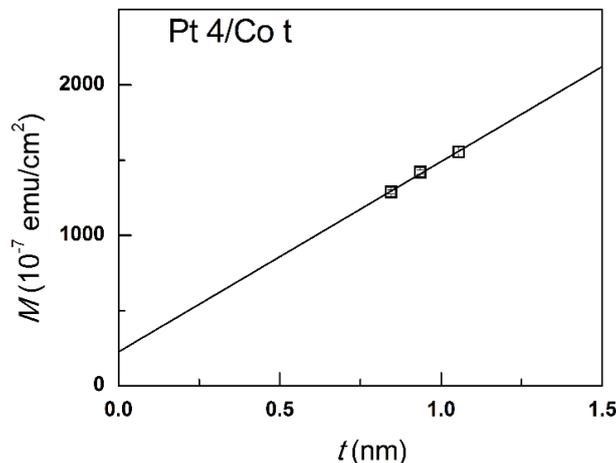

Fig. S6. Magnetic moment per unit area for an as-grown Pt 4/Co t bilayer for three different Co thicknesses.

**III. Resistivity of the heavy metal layers**

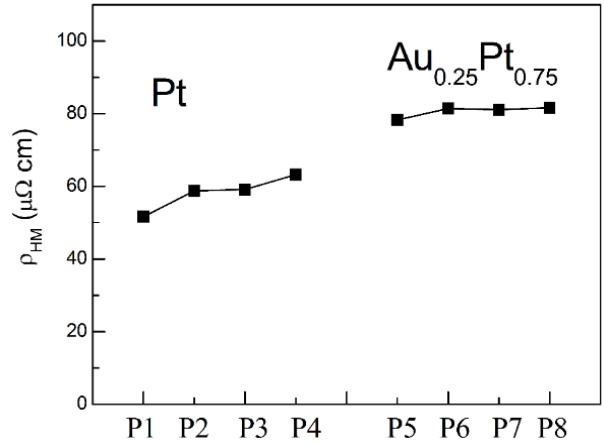

Fig. S7 Resistivity of the heavy metal layers from samples P1-P8 as a function of annealing.

**IV. Enhancement of perpendicular magnetic anisotropy of a Pt 4/Co 1.44 bilayer**

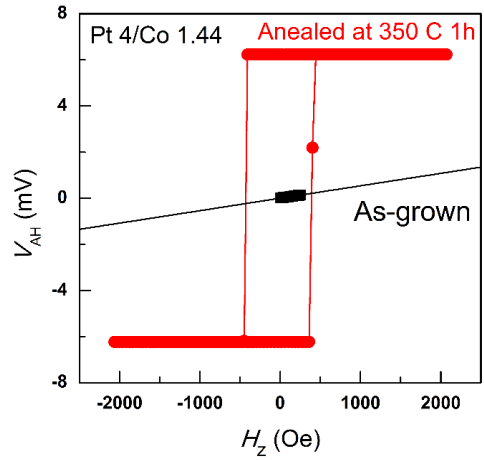

Fig. S8. Enhancement of perpendicular magnetic anisotropy of a Pt 4/Co 1.44 bilayer as indicated by the anomalous Hall voltage hysteresis loops. The increase of the PMA after the first annealing is attributable to the improvement of the sharpness of the Co/Pt interface. It is well known that high temperature annealing can help to improve the magnetic anisotropy properties of Co/Pt superlattices by the separation of Co out from a disordered Co-Pt mixture (see W. Grange et al., Phys. Rev. B **62**, 1157 (2000)).